\documentclass[
    ,final            
    ,numberedheadings 
  ] {aipproc}
\layoutstyle{6x9}
\usepackage{natbib}
\usepackage{textpos}

\def\gr{{$\gamma$-ray}}

\def\nupS{$\nu_{ p}^S$}
\def\nupIC{$\nu_{ p}^{IC}$}
\def\SpIC{$S_{p}^{IC}$}
\def\SpS{$S_{p}^{S}$}
\def\nupSline{$\gamma_{p}^S\propto\nu_{p}^S$}

\newcommand{\lsim}{{\lower.5ex\hbox{$\; \buildrel < \over \sim \;$}}}
\newcommand{\gsim}{{\lower.5ex\hbox{$\; \buildrel > \over \sim \;$}}}

\newcommand{\fermi}{{\it Fermi} }
\newcommand{\frmi}{{\it Fermi}}

\newcommand{\sw}{{\it Swift}}

\begin{document}

\title{The  \textit{Fermi} blazars' divide based on the diagnostic of the SEDs peak frequencies}


\author{A. Tramacere}{
  address={ISDC, Data Centre for Astrophysics, Chemin d'Ecogia 16, CH-1290 Versoix, Switzerland.\\ e-mail:\texttt{andrea.tramacere@unige.ch}}
}

\author{E. Cavazzuti}{
  address={Agenzia Spaziale Italiana (ASI) Science Data Center, I-00044 Frascati (Roma), Italy}
}

\author{P. Giommi}{ 
  address={Agenzia Spaziale Italiana (ASI) Science Data Center, I-00044 Frascati (Roma), Italy}
}
\author{N. Mazziotta}{
 address={Dipartimento di Fisica, ``M.Merlin'' dell' Universit\`a e del Politecnico di Bari, Italy}
  ,altaddress={Istituto Nazionale di Fisica Nucleare, Sezione di Bari, 70126 Bari, Italy } 
}
\author{
C. Monte}{
  address={Dipartimento di Fisica, ``M.Merlin'' dell' Universit\`a e del Politecnico di Bari, Italy}
  ,altaddress={Istituto Nazionale di Fisica Nucleare, Sezione di Bari, 70126 Bari, Italy } 
 } 
\begin{textblock}{15}(2.4,2.4)  
\small{on behalf of the \textit{Fermi}-LAT collaboration }
\end{textblock}

\begin{abstract}
We have studied the quasi-simultaneous Spectral Energy Distributions
(SED) of 48 LBAS blazars, detected within the three months of the  LAT Bright AGN Sample
(LBAS) data taking period, combining Fermi and Swift data with radio NIR-Optical and hard-
X/gamma-ray data.\\

Using these quasi-simultaneous SEDs, sampling both the low  and the high 
energy peak of the blazars broad band emission,  we were able to  apply a diagnostic tool 
based on the estimate of the peak frequencies of the synchrotron (S) and Inverse
Compton (IC) components.\\

Our analysis shows a \fermi blazar's divide based on the peak frequencies of the SED.  
The robust result is that the Synchrotron Self Compton (SSC) region
divides in two the $\nu_{p}^{S}$--$\gamma_{p}^{SSC}$ plane.
Objects within or below this region, radiating likely via the SSC process, are high-frequency-peaked 
BL Lac object (HBL), or low/intermediate-frequency-peaked BL Lac object (LBL/IBL).  
All of the IBLs/LBLs within or below the SSC region are not Compton dominated.\\
The objects lying  above the SSC region, radiating likely via the External
radiation Compton (ERC) process,  are Flat Spectrum Radio Quasars and 
IBLs/LBLs. All of the IBLs/LBLs in the ERC region  show a significant Compton dominance.\\

\end{abstract}

\maketitle

\section{Introduction}

Blazars objects are Active Galactic Nuclei (AGNs) characterized by a polarised 
and highly variable non-thermal  continuum emission extending from radio to 
$\gamma$-rays. In the most accepted scenario, this radiation is produced  
within a relativistic jet that originates in the central engine and points 
close to our line of sight. Since the  relativistic outflow moves with a bulk 
Lorentz factor ($\Gamma$) and is observed at small angles ($\theta\simeq  
1/\Gamma$), the emitted fluxes are affected by a beaming factor $\delta = 
1/(\Gamma (1 - \beta \cos\theta ))$. \\ Despite these objects have been 
observed for more than 40 years over the entire electromagnetic spectrum, 
still present puzzling behaviours. Indeed, their complex 
multiwavelength variability requires the use of  simultaneous multi-frequency 
data to fully understand the underling physical scenario.\\ The study of the 
Spectral Energy Distribution (SED) of blazars has been largely enriched  since 
July 2008 after the beginning of the scientific activity of the $\gamma$-ray 
Large Area Telescope (LAT)  \cite{Atwood2009} on board \textit{Fermi}-GST 
\citep{Ritz2007}.\\ A detailed investigation of the broad-band spectral 
properties of 48 $\gamma$-ray selected blazars observed by  Fermi is reported 
in \cite{SEDpaper}. The quasi-simultaneous SEDs presented in that paper were 
obtained using data from \frmi~ operating in survey  mode, combined to 
simultaneous data from \sw, and other high-energy astrophysics satellites 
mission on orbit,  complemented by other space and ground-based 
observatories.\\ Here we focus on the physical implications of these SEDs, 
using the peak frequencies of the  synchrotron (S) and inverse Compton (IC) 
component as a diagnostic tool to discriminate among different emission  
scenarios.

\section{Implications for physical modeling}
\label{sec:sedsparameters}
The quasi-simultaneous SEDs reported in \cite{SEDpaper,Monte,Gaspa} show the 
typical two bump shape that is seen in radio  or X-ray selected blazars. 
According to current models the low energy bump is interpreted as synchrotron  
emission from highly relativistic electrons, and the high energy bump is 
related to inverse Compton emission of various underlying radiation fields.\\ 
In the case of synchrotron self Compton model (SSC)
\citep{Jones74,Ghisellini1989} the seed photons for the IC process are 
the synchrotron photons produced by the same population of relativistic 
electrons.\\ In the case of external radiation Compton (ERC) scenario 
\citep{Sikora94,Dermer02}, the seed photons for the IC process are typically UV 
photons generated by the accretion disk surrounding the black hole, and 
reflected toward the jet by the Broad Line Region (BLR) within a typical 
distance from the accretion disk of the order of one pc. If the emission occurs 
at larger distances, the external radiation is likely to be provided by a dusty 
torus (DT) \citep{Sikora02}. In this case the photon field is typically peaked 
at IR frequencies.\\ 

Blazars come in two main flavours: BL Lac objects  and Flat Spectrum Radio 
Quasars (FSRQ). The former type is  characterised by featureless optical 
spectra and their SEDs are usually interpreted in the framework of pure SSC  
scenario. On the contrary, FSRQs display the prominent emission lines that are 
typical of QSOs, and are likely to  have the IC component dominated by the ERC 
emission. BL Lac objects are often subdivided into three subclasses  depending 
on their SEDs. This classification \cite{Padovani1995} uses the peak energy of 
the synchrotron emission,  which reflects the maximum energy the particles can 
be accelerated in the jet, to classify BL Lacs into low energy,  intermediate 
energy and high energy synchrotron peak objects, respectively called LBL, IBL 
and HBL. \\ 

We follow a phenomenological approach to obtain information about the peak 
Lorentz factor of the electron distribution ($\gamma_{p}$) most contributing 
to the synchrotron emission and to the inverse Compton process. To test the 
methods used to estimate $\gamma_{p}$, we employ an accurate  numerical 
model \citep{Tramacere09,Tramacere07a,Massaro06,Tramacere03} that can reproduce 
both the SSC and ERC models. For the electron distribution we considered a 
log-parabola of the form $n(\gamma)=K\cdot10^{~r~Log(\gamma/\gamma_{p})^2}$ 
with $\gamma_{p}$ ranging between 100 and $6\cdot10^5$ and with curvature 
parameter $r=2.0$ \citep{Massaro04,Tramacere07b}. As input parameter for the 
benchmark SSC model we use a source size $R=10^{15}$ cm, a magnetic field 
$B=0.1$ G, a beaming factor $\delta=10$, and an electron density N=1 
$e^-/cm^{3}$ (N=$\int n(\gamma)d\gamma$). In the case of the benchmark ERC 
model, we use the same set of parameters with the addition of the external 
photon field produced by the accretion disk and reflected by the BLR toward the 
emitting region with an efficiency $\tau_{BLR}=0.1$. The accretion disk 
radiation is modelled by a multitemperature black body, with a innermost disk 
temperature of $10^5$ K. 

\section{The synchrotron peak frequency}
The dependence of the observed peak frequency of the synchrotron emission 
(\nupS) on magnetic field intensity ($B$), electron Lorentz factor ($\gamma$), 
beaming factor ($\delta$) and redshift ($z$) is  given by: 

\begin{equation} 
\nu_{p}^{S}=3.2\times10^6(\gamma^S_{p})^2 B \delta/(1+z)=\nu_{p}^{S'}\delta/(1+z). 
\label{nupS} 
\end{equation} 

where $\nu_{p}^{S'}$ is the synchrotron peak frequency in the emitting region 
rest-frame. A good estimate of $\gamma^S_{p}$ in terms of the differential 
electron energy distribution ($n(\gamma)=dN(\gamma)/d\gamma$) is given by the 
peak of $\gamma^3 n(\gamma)$, hereafter $\gamma_{3p}$ 
\citep{Tramacere09,Tramacere07b}. 
The value of $\gamma_{p}^S$ is estimated by fitting the peak of the numerically 
computed synchrotron SED with a log-parabolic analytical function. Note, however, 
that there is a degeneracy on the value of $\gamma_{p}^S$ given by the product 
$B\delta$. We discuss this point in the next sections. 

\begin{figure}
\begin{tabular}{lr}
\includegraphics[width=9.7cm,angle=0]{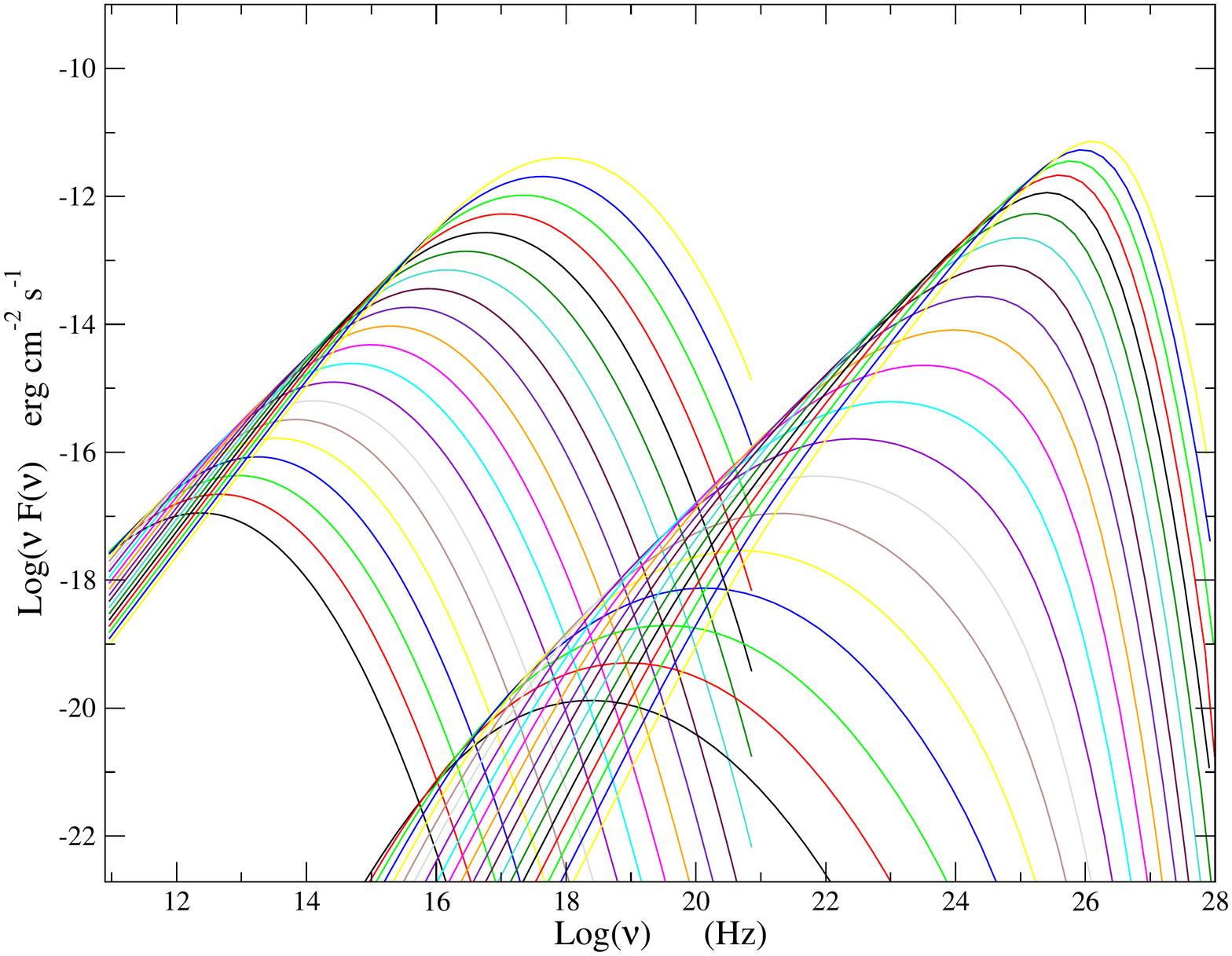} & \includegraphics[width=7.4 cm,angle=0]{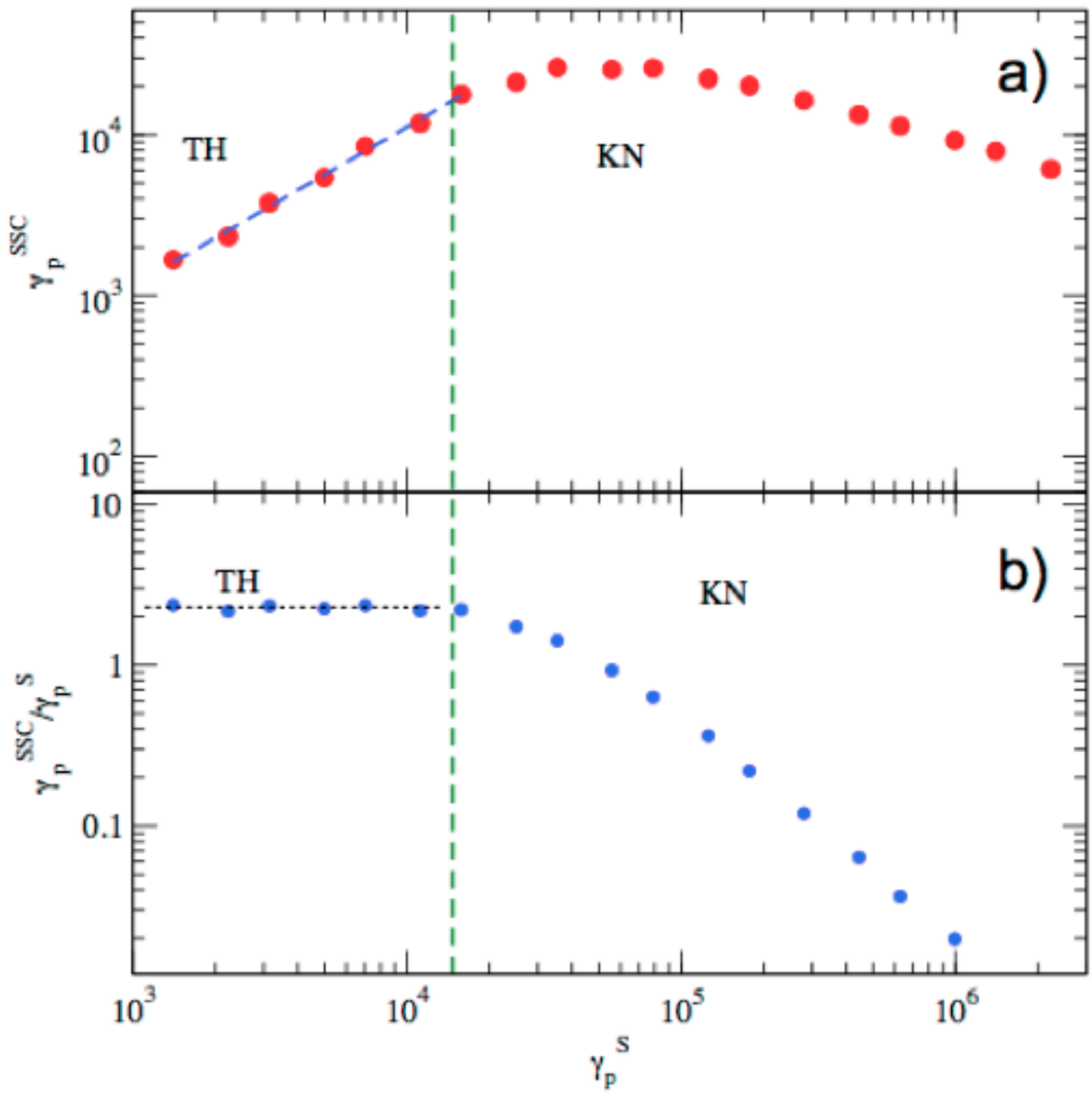}\\ 
\end{tabular}
\caption{\textit{ Left panel:} SSC SEDs obtained using as electron distribution 
a log-parabola $n(\gamma)=K\cdot10^{~r~Log(\gamma/\gamma_{p})^2}$ with 
$\gamma_{p}$ ranging between 100 and $6\cdot10^5$, and the curvature 
parameter $r=2.0$. The other model parameters are: source size $R=10^{15}$ cm, 
a magnetic field $B=0.1$ G, a beaming factor $\delta=10$, and an electron 
density N=1 $e^-/cm^{3}$ (N=$\int n(\gamma)d\gamma$).  ~~\textit{Right panels:} 
Estimate of $\gamma_{p}^{S}$ and $\gamma_{p}^{SSC}$ for  numerically 
computed SEDs showed in the left panel From to top to bottom: \textit{ a)} 
$\gamma_{p}^{SSC}$ as a  function of $\gamma_{p}^S$, the transition from 
the TH trend (blue dashed line) to the KN region is evident for $\gamma>2\cdot 
10^4$. \textit{b)}The ratio of $\gamma_{p}^{SSC}$ to $\gamma_{p}^S$, also 
in this case, above the TH region (vertical dashed green line) it is evident the 
effect of the KN suppression, $\gamma_{p}^{SSC}$ gets to increasingly 
underestimate $\gamma_{p}^S$ as $\gamma_{p}^S$ is increasing.} 
\label{fig:KN-STUDY} 
\end{figure} 

\section{The inverse Compton peak frequency}
In a simple SSC model, and under the Thomson regime (TH) of the IC scattering, 
the observed peak frequency of the  synchrotron component (\nupS) is related to 
the observed peak frequency of the inverse Compton one (\nupIC) by the  
following relation: 

\begin{equation} {\nu_{p}^{IC} \over \nu_{p}^{S}} \simeq \frac{4}{3}(\gamma^{SSC}_{p})^2 
\label{nupSSC} 
\end{equation} 

where $\gamma_{p}^{SSC}$ is of the same order of $\gamma^S_{p}$ and of
$\gamma_{3p}$. In the left 
panel of of Fig. \ref{fig:KN-STUDY} we plot the S and IC component for a SSC 
scenario for the choice of SSC parameters  reported in Sect. 2. In the right panels 
(\textit{a} and \textit{b}) of the same figure we show that trend predicted  
by Eq. \ref{nupSSC} is valid only for $\gamma_{p}^{SSC} \lsim 2\cdot 10^4$ 
where the transition from  Thomson to Klein Nishina (KN) regime occurs. In the 
KN regime Eq. (\ref{nupSSC}) is no longer valid: in fact, the  kinematic limit 
for the maximum energy of the up-scattered photons in the emitting region 
rest-frame is: \begin{equation} \nu_{max}^{IC} = 
\frac{4\gamma^2\nu_S}{1+4\gamma^2(h\nu_S/m_ec^2)}~~. \end{equation} 

As the energy of the seed photons in the electron rest-frames increases, the
maximum up-scattered photon energies approaches the energy of the up-scattering
electron ($\gamma m_ec^2$). This means that the peak energy of the IC emission
is no longer growing with $(\gamma_{p}^{SSC})^2$ according to Eq. (\ref{nupSSC}), 
but it starts becoming smaller as shown in panels \textit{a} and \textit{b} of 
Fig. \ref{fig:KN-STUDY}. Above the TH region (vertical dashed green line), it is 
evident the effect of the KN suppression, and $\gamma_{p}^{SSC}$ gets to 
increasingly underestimate $\gamma_{p}^S$ as $\gamma_{p}^S$ is increasing.
We note that this effect is particularly relevant for the case of HBL objects.\\ 

Other deviations from the trend predicted by Eq. 
(\ref{nupSSC}) occur when further radiative components add to a single zone 
SSC. In fact, for the case of External Compton scenario, the observed peak 
frequency  of the ERC component in terms of the frequency of the external 
photon field in the disk rest-frame ($\nu_{p}^{' EXT}$) reads: 

\begin{equation} 
\frac{\nu_{p}^{ERC} }{ \nu_{p}^{' EXT}\Gamma} \simeq (\frac{4}{3}) (\gamma^{ERC}_{p})^2\delta/(1+z) 
\label{nupERC} 
\end{equation} 

where $\nu_{p}^{' EXT}\Gamma$ is the external photon field frequency 
transformed to the rest-frame of the emitting region which is moving with a 
bulk Lorentz factor $\Gamma$, and assuming that the BLR radiation is isotropic.

If one uses Eq. (\ref{nupSSC}) in place of Eq. (\ref{nupERC}) (an assumption 
justified by the fact that the UV and IR external radiation fields are usually 
dominated by the non-thermal synchrotron emission of the source), a significant 
bias on the value of $\gamma^{ERC}_{p}$ is introduced. 
In fact, the resulting value of $\gamma_{p}$ is strongly overestimated in 
the case of external UV radiation field 
($\gamma_{p}^{SSC}>>\gamma^{ERC}_{p}$ and 
$\gamma_{p}^{SSC}>>\gamma^{S}_{p}$ ). In the case of IR  external 
radiation field, the bias is smaller but the measured value of 
$\gamma^{SSC}_{p}$ is still  overestimating both $\gamma^{ERC}_{p}$ and 
$\gamma^{S}_{p}$. 

\begin{figure}
\begin{tabular}{ll}
\includegraphics[width=8.cm,angle=0]{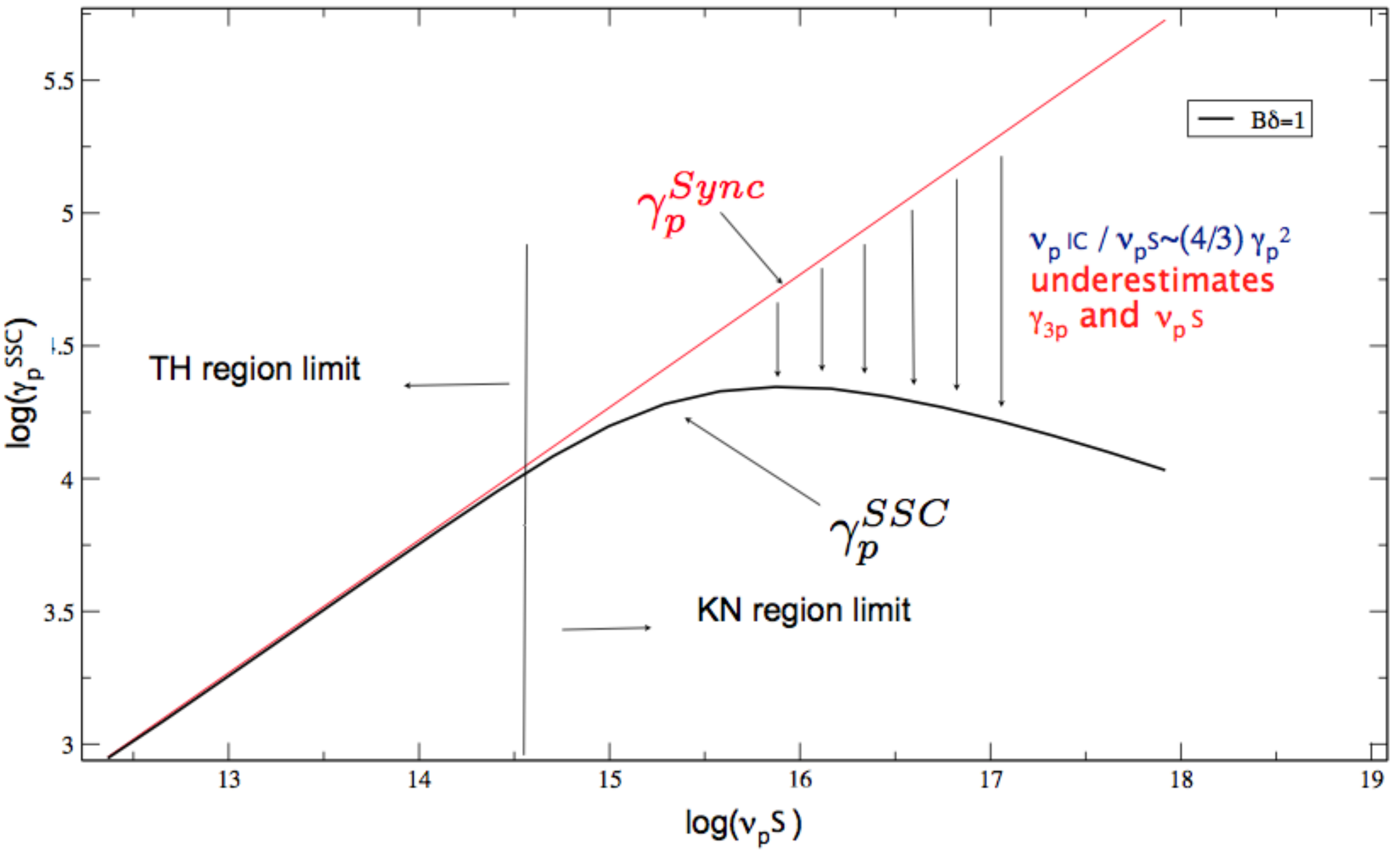}&\includegraphics[width=8.0cm,angle=0]{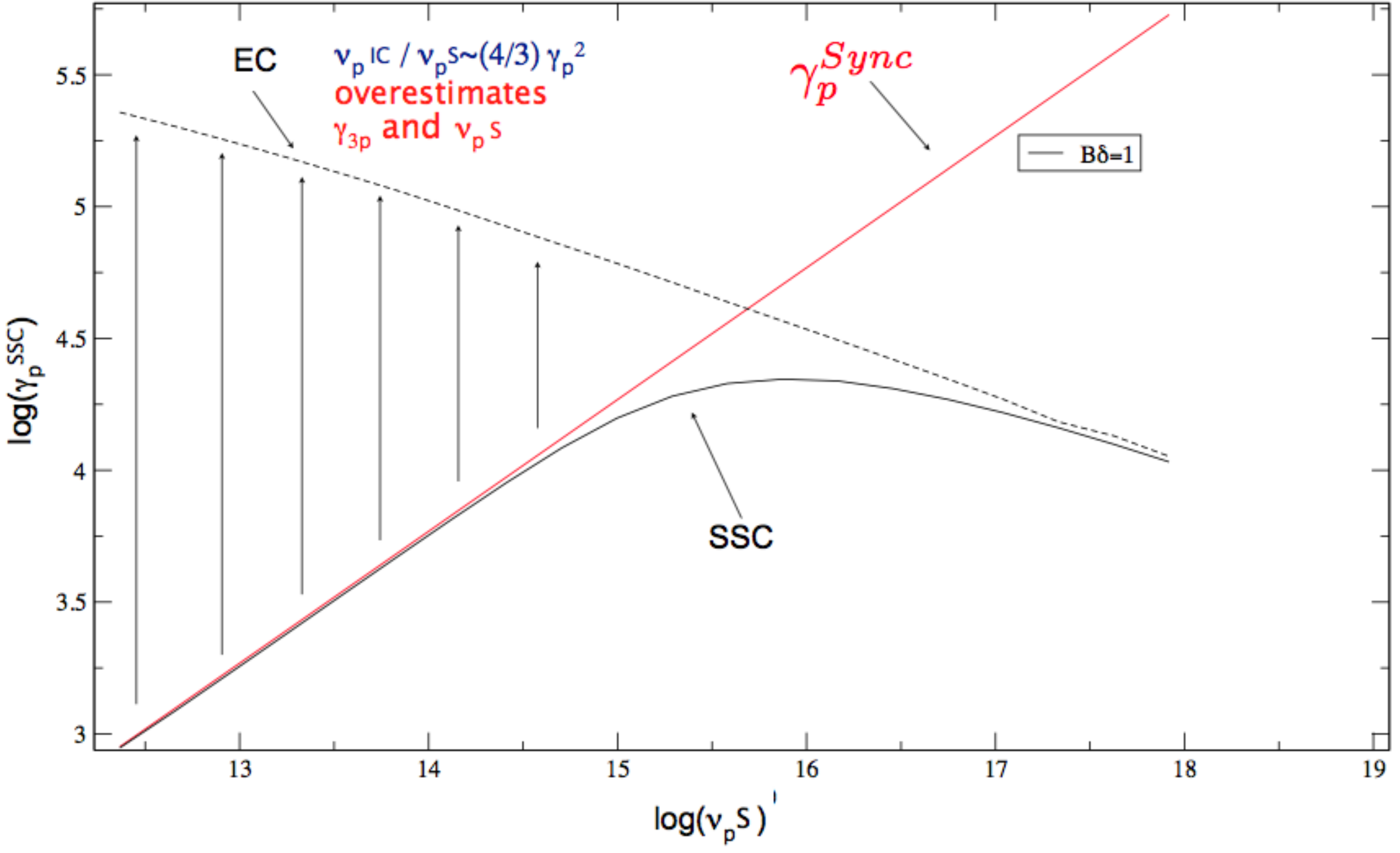}\\ 
\caption{ \textit{Lef Panel:} graphical representation of the deviation from the 
trend in Eq. \ref{nupSSC} in the case of SSC emission in KN regime. 
\textit{Right panel:} graphical representation of the deviation from the  trend 
in Eq. \ref{nupSSC} in thecase of ERC emission. } 
\label{fig:SSC-EC-BIAS} 
\end{tabular} 
\end{figure}

In conclusion, when $\gamma_{p}$ is estimated through Eq. (\ref{nupSSC}) we 
expect two main biases: 

\begin{enumerate}

\item a bias related to the KN effect, affecting mostly objects radiating via SSC
under the KN regime, 
which leads to an underestimate of $\gamma_{p}^S$ and $\gamma_{3p}$. This trend 
is reported in the left panel of Fig. \ref{fig:SSC-EC-BIAS}. 

\item A bias related to the ERC  scenario, which yields an overestimate of 
 $\gamma_{p}^S$ and $\gamma_{3p}$. This trend is 
reported in the right panel of Fig. \ref{fig:SSC-EC-BIAS}. 

\end{enumerate}

\section{Application of  the peak diagnostic tool to the LBAS MW SEDs}
\begin{figure}
\includegraphics[width=12.cm,angle=0]{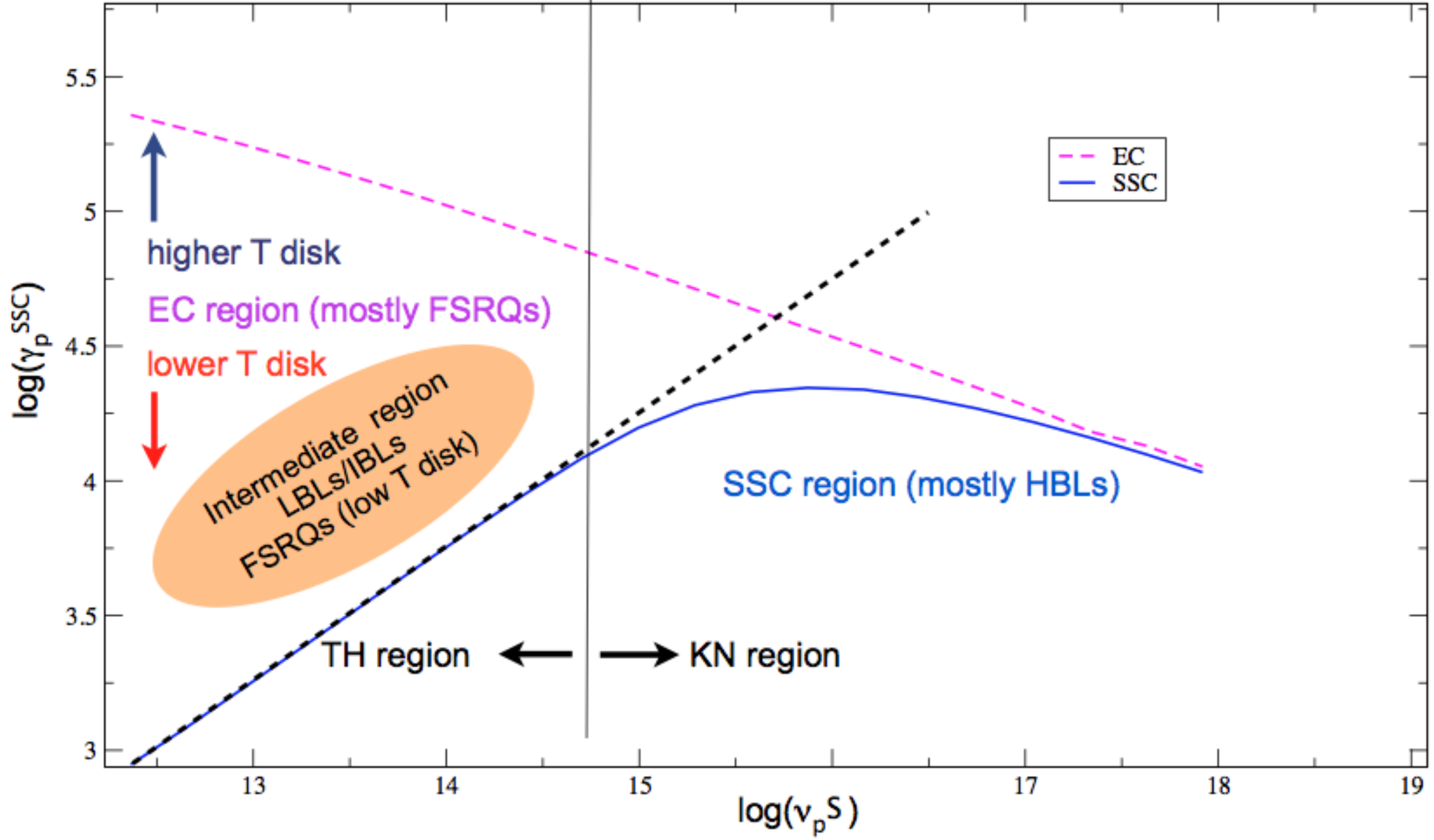}
\caption{Schematic representation of the divide predicted
by the SSC/KN and ECR deviations from the Eq. \ref{nupSSC} trend. 
The dashed black line represent the \nupSline~ trend, the  solid blue 
line represents the SSC trend from the TH to the KN regime, and the dashed
purple line represents the ERC trend.}
\label{fig:scheme}
\end{figure}

The arguments presented in the previous section provide an interesting diagnostic 
tool in the $\nu_{p}^{S}$--$\gamma_{p}^{SSC}$ plane. 
Since under the TH regime $\gamma_p^{S}\simeq\gamma_p^{SSC}$, we expect objects radiating 
\gr s mainly via the SSC/TH mechanism to lay along the \nupSline~ line, 
and below it in the case of SSC/KN regime. Objects radiating mainly via 
the ERC mechanism are expected to lay above  the \nupSline~ line.
A schematic expectation for this divide is plotted in Fig. \ref{fig:scheme},
where the dashed black line represents the \nupSline~ trend, the  solid blue 
line represents the SSC trend from the TH to the KN regime, and the dashed
purple line represents the ERC trend.
\\ 

As a caveat, we remind the reader that the peak frequencies derived
in \cite{SEDpaper} may have systematics. Indeed, for sources peaking
below (MeV blazars or LBL) or above (HBLs) the \fermi window, the
peak frequency extrapolated from the polynomial fit \cite[see][]{SEDpaper,Gaspa,Monte},
may have an uncertainty  up to a factor of two. A more detailed 
analysis that takes this effect into account will be presented in a further 
paper. We stress that this bias can affect only these single sources
and  the understanding of some outliers, and does not
affect the overall trend that is the goal of the present analysis. \\

\begin{figure}
\includegraphics[width=16.cm,angle=0]{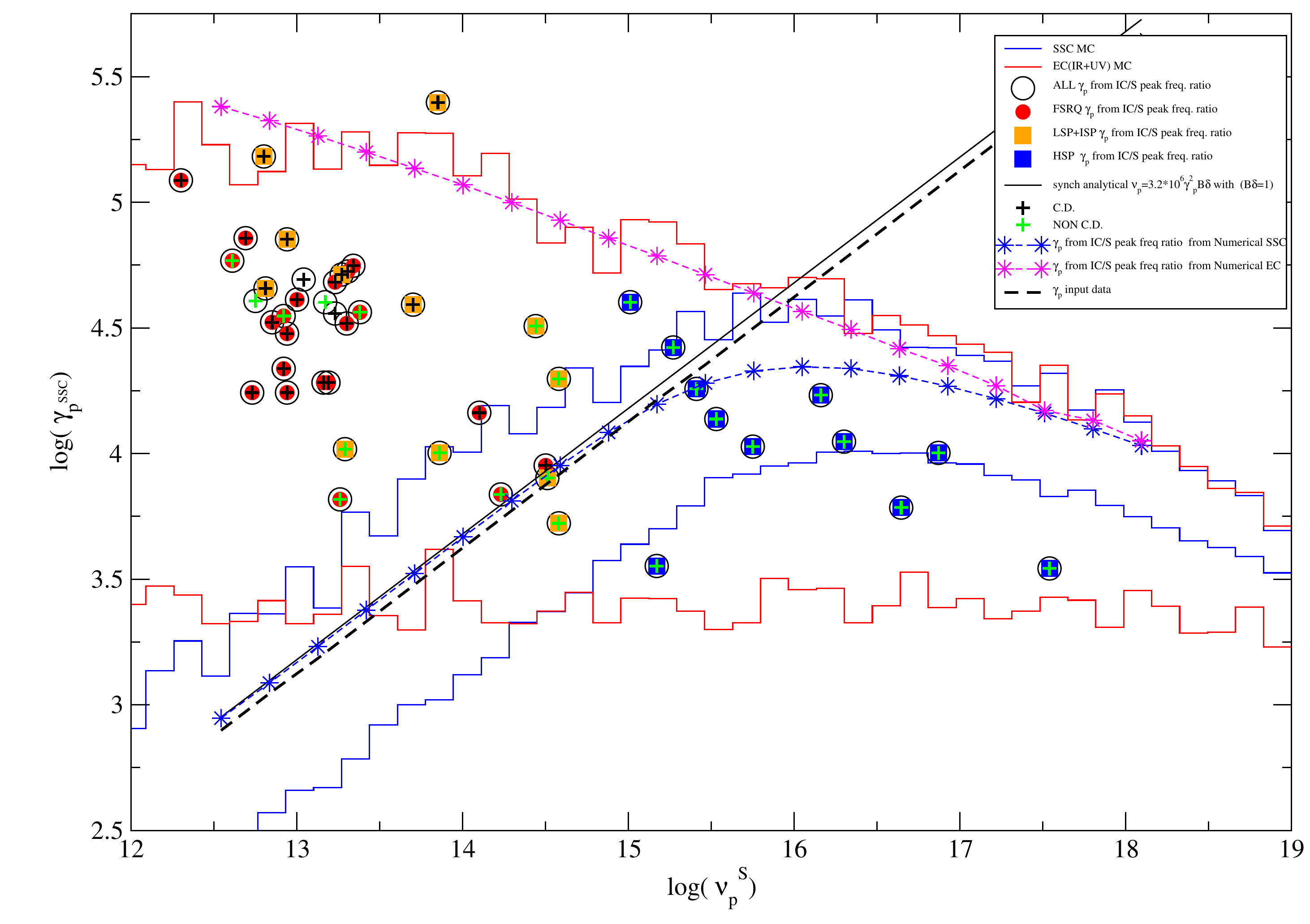}
\caption{ $\gamma_{p}^{SSC}$ obtained by Eq. \ref{nupSSC}  for
the objects reported in Table 10. blue solid boxes represent HPB
objects, orange solid boxes represent IPBs/LPBs objects and red
solid circles represent FSRQs. The black solid line represents
$\nu_{p}^S$ estimated by Eq. (\ref{nupS}) for both the ERC and
the SSC numerical SEDs, the blue solid line represents
$\gamma_{p}^{SSC}$ estimated from Eq. (\ref{nupSSC}) applied to
numerically computed the SSC SEDs, and the solid purple line
represents the same for the case of ERC emission. The true value
of the simulation  is represented by the black dashed line.
Parameters of the model are given in Sect. 9. The blue and red
contours delimit the area covered by the estimate of
$\gamma_{p}^{SSC}$ for the case of SSC and ERC models
respectively and for a Monte Carlo simulation with  values of
$\delta$ ranging between 10 and 15,  $B$  ranging between  0.01
and 1G and $T$ ranging between 10 and 10$^{4.5}$ K. }
\label{fig:nupvgammael}
\end{figure}

To test this scenario we use the values of $\gamma_{peak}^{SSC}$ obtained by Eq. 
(\ref{nupSSC}) applied to the numerically computed SSC/ERC SEDs , and we 
compare these trends with those obtained applying Eq. (\ref{nupSSC}) to the
observed peak frequencies and fluxes reported in the Tab. 13 of \cite{SEDpaper}.\\
Fig. \ref{fig:nupvgammael} shows the  location of HBL objects (blue solid boxes), 
IBLs/LBLs objects (orange solid boxes) and FSRQs (red solid circles).\\ 
The values of $\gamma_{peak}^{SSC}$ estimated for the case of SSC emission 
(dashed blue line with stars) show clearly the effect of the transition from 
the TH to the KN regime. 
We note that all but two of the HBLs lay below the \nupSline~ line. 
In particular all the HBL objects below the \nupSline~ line have $\gamma_{p}^{SSC}$ 
values below the prediction of the SSC scenario (solid blue line), confirming, 
as expected,that the SSC emission occurs under the KN regime.\\
On the contrary, all the FSRQs and the LBL/IBL objects but one lay above the 
 \nupSline~ line. All the FSRQs objects but one 
have a value of $\gamma_{p}^{SSC	}$ in excess of a factor $\sim 10^4$ and limited 
by the prediction from the ERC model (purple dashed line with stars).\\ 
The LBLs/IBLs sources are more uniformly distributed across the region delimited 
by the the SSC TH prediction and by the ERC one.\\
By further dividing the sample in Compton Dominated (CD) objects 
( \SpIC $>$2 \SpS) \footnote{We indicate $\nu_{p}F(\nu_{p})$ as $S_p$} 
and non-Compton Dominated (NCD) objects (\SpIC $\le$ 2 \SpS), we found that all 
the CD objects lay above the \nupSline~ line and 
populate the region between  the SSC TH and the ERC regime, with the FSRQs 
clustering toward the ERC region.\\

Our analysis shows that the ERC model could explain the
high CD values as well as the high values of $\gamma_{p}^{SSC}$
estimated in the case of FSRQs and IBLs/LBLs.
In order to explain the high values of $\gamma_{p}^{SSC}$
obtained in the case of FSRQs in the context of single
zone SSC emission model, a very small value of the magnetic field with
($B<0.01$ G) is required.\\

\subsection{$B\delta$ degeneracy and Monte Carlo approach}
As a final step, we discuss two additional effects that have consequences for 
the source distribution is this  parameter space: 

\begin{enumerate} \item The $B\delta$ degeneracy on $\gamma_{p}^{S}$ 
can affect the transition region from TH to KN regime, since high values of 
$\delta$ allow the TH regime to propagate towards higher frequencies. \item The 
values of $\gamma_{p}$ in the case of an UV external radiation field (purple 
line Fig.\ref{fig:nupvgammael} ) constitutes an upper limit to the observed 
values of $\gamma_{p}$, meaning that objects in the region below the ERC 
prediction line require a wider range of external photon energies, extending 
down to the IR band. \end{enumerate}

To take into account both these effects we perform Monte Carlo (MC) simulations.
Specifically, we generate both the SSC and ERC numerical computation of the 
SEDs extracting $\delta$, $B$ and the temperature of the accretion disk $T$ 
from a random uniform distribution, in order to cover a larger volume of the 
parameter space.  
We generate 1000 realizations, with 
$\delta$ ranging in the interval [10-15], $B$ in the interval [0.01-1] G and 
$T$ in the interval [$10-10^{4.5}$] K. In Fig. \ref{fig:nupvgammael} the MC 
results for the case of SSC lay within the area delimited by the blue  contour 
line, while the results in the case of ERC model, are delimited by the light 
red contour line. \\ 

When we compare the observed data with the MC results, we note the following:

\begin{enumerate}
\item The MC simulations, compared to the ERC one for the only case of 
UV external photons (purple line),  cover a much wider region of the parameter 
space. In the case of the MC SEDs, the range of temperatures of the BB  
emission allows us to take into account external photon fields peaking at IR 
frequencies. The FSRQs and IBLs/LBLs populate the whole parameter space 
delimited by the ERC/UV (purple line) and the SSC/TH case (solid  blue line, 
below about $10^{15}$ Hz). This suggests, that in the ERC paradigm, the 
observed data concerning FSRQs (red  circles) and IBLs/LBLs (orange square symbols), 
require external photon fields ranging form the UV down to the IR.\\
\item In the case of the MC realization, all the HBLs and all the non-Compton 
dominated LBLs/IBLs  are compatible with the SSC 
prediction, both in TH and KN regime.\\ 
\item All the FSRQs but four, are not compatible with the MC SSC
region. Two of the FSRQs consistent with the MC SSC region, are non-CD.\\ 
\item All the LBLs/IBLs that are CD, are not consistent with the MC SSC.
\item the empty circles in Fig. \ref{fig:nupvgammael}, are blazars of unknown
type. 
\end{enumerate}

\section{Discussion and Conclusions}
Trough the construction of quasi-simultaneous SEDs, sampling both the low 
and the high energy peak of the blazars broad band emission, we were able to 
apply a diagnostic tool based on the ratio of the peak frequencies. 
We studied  the deviations from the trend predicted by Eq. 
\ref{nupSSC}, given by the KN regime (relevant to the case of HBLs), and by the 
ERC emission (relevant to the case of FSRQs), we were able to discriminate 
among different emission scenarios.\\

Our analysis shows that the HBL objects attend the SSC prediction, and 
that the values of $\gamma_{p}^{SSC}$ 
returned for sources peaking above about $ 10^{15} $ Hz are within the KN limit 
expectation. All of the HBLs are not CD, as expected in the case of SSC 
emission.\\ 
FSRQs populate mainly the ERC region of Fig. \ref{fig:nupvgammael}, and 
the dispersion on the values of $\gamma_{p}^{SSC}$ hints for a 
dispersion in the typical temperature of the BB emission. 
This dispersion on the temperature is consistent with the dominant external photon 
field originating from some sources in the BLR, and in the DT for others.\\ 
All of the FSRQs but 5, are Compton Dominated. Two of the non-CD FSRQs lay close 
the SSC/TH region.\\ 
The LBL/IBL objects cover the region of Fig. \ref{fig:nupvgammael} ranging from the 
SSC/TH to the ERC/BLR. The non-CD LBLs/IBLs are consistent with the SSC/TH region. 
On the contrary, the CD LBLs/IBLs cluster  in the ERC/BLR region. This feature is 
very interesting since many BL Lac objects, and BL Lacertae itself, show 
intermittent emission lines. At this regard a monitoring of the optical 
spectrum compared the \gr flaring state and Compton dominance could be very 
useful to confirm  the ERC component also for this class of objects. \\

In conclusion, our analysis shows a \fermi blazar's divide, based on the
peak frequencies of the SED. The robust result is that the SSC region
divides in two the $\nu_{p}^{S}$--$\gamma_{p}^{SSC}$ plane.
Objects within or below this region, radiating likely via the SSC
process are HBLs, and IBLs/LBLs without Compton dominance.\\
The objects lying  above the SSC region, radiating likely via the 
ECR process,  are FSRQs and IBLs/LBLs and are mostly Compton dominated.


\begin{theacknowledgments}
The \fermi LAT Collaboration acknowledges the generous
support of a number of agencies and institutes that have supported the $Fermi$
LAT Collaboration. These include the National Aeronautics and Space
Administration and the Department of Energy in the United States, the
Commissariat \`a l'Energie Atomique and the Centre National de la Recherche
Scientifique / Institut National de Physique Nucl\'eaire et de Physique des
Particules in France, the Agenzia Spaziale Italiana and the Istituto Nazionale
di Fisica Nucleare in Italy, the Ministry of Education, Culture, Sports,
Science and Technology (MEXT), High Energy Accelerator Research Organization
(KEK) and Japan Aerospace Exploration Agency (JAXA) in Japan, and the K.\ A.\
Wallenberg Foundation, the Swedish Research Council and the Swedish National
Space Board in Sweden.
Additional support for science analysis during the operations phase from the 
following agencies is also gratefully acknowledged: the Istituto Nazionale di 
Astrofisica in Italy and the K.~A. Wallenberg Foundation in Sweden. This 
research is based also on observations with the 100-m telescope of the MPIfR 
(Max-Planck-Institut f\"ur Radioastronomie) at Effelsberg. 
\end{theacknowledgments}

\bibliographystyle{aipproc}   

\bibliography{tramacere}

\end{document}